\documentclass[9pt,twocolumn,twoside]{opticajnl}
\journal{opticajournal} % use for journal or Optica Open submissions

\setboolean{shortarticle}{true}
\usepackage{lineno}
\title{Non-Hermitian dynamical topological winding in photonic mesh lattices}

\author[1,2,*]{Stefano Longhi}

\affil[1]{Dipartimento di Fisica, Politecnico di Milano, Piazza L. da Vinci 32, I-20133 Milano, Italy}
\affil[2]{IFISC (UIB-CSIC), Instituto de Fisica Interdisciplinar y Sistemas Complejos - Palma de Mallorca, Spain}

\affil[*]{stefano.longhi@polimi.it}

\begin{abstract}
Topological winding in non-Hermitian systems are generally associated to the Bloch band properties of lattice Hamiltonians. However, in certain non-Hermitian models topological winding naturally arise from 
the dynamical evolution of the system and related to a new form of geometric phase. Here we investigate dynamical topological winding in non-Hermitian photonic mesh lattices, where the mean survival  time of an optical pulse circulating in coupled fiber loops is quantized and robust against Hamiltonian deformations. The suggested photonic model could provide an experimentally accessible platform for the observation of non-Hermitian dynamical topological windings.
\end{abstract}

\setboolean{displaycopyright}{false} % Do not include copyright or licensing information in submission.

\begin{document}

\maketitle

%{\em Introduction.} 

{\em Introduction.} In the past few decades, an explosion of interest has been devoted to topological phenomena in several areas of physics, ranging from condensed matter physics \cite{r1,r2} to photonics \cite{r3,r4,r5,r6}. Recently, the concept of topological phases has been extended to non-Hermitian (NH) systems \cite{r7,r8,r9,r10,r11,r12,r13,r14,r15,r16,r17,r18,r19}, where the eigenenergies of the underlying Hamiltonian can be complex. Special attention has been paid to the Bloch band spectral topology of NH lattices and its connection to the NH skin effect \cite{r11,r12,r13,r14,r17,r18}. 
Dynamical methods for detecting winding numbers in both Hermitian and NH lattice models have been suggested and experimentally demonstrated on several occasions \cite{r19,r20,r21,r22,r23,r24,r25,r26,r27,r28,r29}. Such winding numbers arise from the underlying Bloch band properties of lattice systems, possibly in the presence of disorder \cite{r7,r30,r31,r32,r33}, however an interesting question is whether other types of topological numbers can be found, which arise from the pure NH dynamical evolution of a system without any spectral origin nor related to lattice structure.  Recently, a new kind of geometric phase has been introduced \cite{r34} to describe the mean return time of periodically-measured quantum systems in finite-dimensional Hilbert spaces \cite{r34,r35,r36,r37,r38}. In such systems, the mean return time can be expressed in terms of a winding number $w$, which equals the number of distinct energy phases of the Hermitian Hamiltonian touched by the dynamics \cite{r34,r36}, and thus rapidly increasing with the system size.  The experimental observation of such dynamical topological winding remains so far elusive in truly quantum systems, mainly because of the long observation times required to resolve the energy levels and unambiguously detect the topological invariant \cite{r38}. Since repeated projective quantum measurements effectively introduce a non-Hermitian dynamics \cite{r23,r35}, such a geometric phase can be regarded as universal one to a certain class of NH systems, i.e. not necessarily of quantum nature. Interestingly, dynamics of classical light in coupled waveguide systems can emulate quantum projective measurements and wave function collapse, and has been previously harnessed to emulate quantum Zeno dynamics in classical settings \cite{r38b,r38c,r38d}.\\ 
In this Letter dynamical topological windings are introduced in photonic systems at the classical level, namely  in NH photonic mesh lattices describing the dynamics of optical pulses in coupled fiber loops \cite{r32,r39,r40,r41,r42}. 
In such systems,  the mean survival  time of an optical pulse circulating in the fiber loops is quantized, and its value is robust against Hamiltonian deformations that avoid energy degeneracies. The present results suggest that  photonics could provide an experimentally accessible platform for the observation of NH dynamical topological windings beyond the NH Bloch band paradigm, and could stimulate further investigations with potential interest in optically-oriented applications, such as in advanced optical sensing.\\
\par
{\em Topological winding in NH dynamics.}  In order to introduce the concept of dynamical NH topological winding \cite{r34,r36,r38}, let us consider a rather arbitrary network comprising $L$ nodes $|n \rangle$ ($n=1,2,...,L$) and described by the single-particle Hermitian Hamiltonian $
{\hat H}=\sum_{n,l} H_
{n,l} |n \rangle \langle l|$. Let us assume that at initial time $t=0$ the system is excited in the node $|q \rangle$, i.e. $|\psi(t=0) \rangle=|q \rangle$ for the vector state $| \psi(t) \rangle$. In a time interval $\tau$, the free evolution of the system is described by the unitary operator ${\hat U}=\exp(-i \tau \hat{H})$.  Indicating by $\hat{\Gamma}= | p \rangle \langle p |$ the projection operator for node $|p\rangle$, we introduce non-Hermiticity in the system via repeated projective measurements every time interval $\tau$ to node $| p \rangle$ \cite{r23,r34,r35,r36}. This is equivalent to consider  a NH system described by the effective NH Hamiltonian $\hat{H}_{eff}$ via the relation \cite{r23,r35}
\begin{equation}
\exp(-i \hat{H}_{eff} \tau)=  \hat{U} (1-\hat{\Gamma}).
\end{equation}
The stroboscopic NH evolution of the wave functioin $|\psi_m \rangle= | \psi(t_m) \rangle $ at times $t=m \tau^-$ $(m=1,2,3,...$) is thus given by $|\psi_{m+1} \rangle=  \hat{U} (1-\hat{\Gamma}) |\psi_{m} \rangle$  with $| \psi_1 \rangle=\hat{U} |q \rangle$. After letting $\theta_m= \langle p | \psi_m \rangle$,
in the periodically-monitored quantum system the quantity $P_m=|\theta_m|^2$ gives the probability that the particle is detected (and thus destroyed) at time $t_m= m \tau$ by a detector placed at site $|p \rangle$ \cite{r34,r36}. The quantum system is said to be recurrent whenever $\sum_{m=1}^{\infty} P_m=1$, i.e. whenever with certainty the particle is destroyed at some observation time \cite{r34}. In this case  the mean survival time in units of $\tau$ is given by
\begin{equation}  
\langle m \rangle = \sum_{m=1}^{\infty} m P_m.
\end{equation}
This time also corresponds to the particle 'transition time' from $|q \rangle$ to $|p \rangle$, or to the 'return time' when $p=q$.
A main central result, which was proven in Ref.\cite{r34}, is that when $p=q$, i.e. when we detect the return time of the quantum particle to its initial state and thus a closed cycle is realized,  the system is always recurrent and the mean survival time $\langle m \rangle$ is quantized. Namely,  $\langle m \rangle$ can be expressed in terms of the winding $w$  \cite{r34,r36,r38}
\begin{equation}
\langle m \rangle = w= \frac{1}{2 \pi i } \int_{0}^{2 \pi} d \varphi \frac{d}{d \varphi} \log \Theta (\varphi) = \frac{1}{2 \pi} \int_{0}^{2 \pi} d \varphi \frac{d \omega}{d \varphi}
\end{equation}
where $\Theta(\varphi)=\sum_{m} \theta_m  \exp(i m \varphi)$ is the generating function of the detection amplitudes $\theta_m= \langle p | \psi_m \rangle$ at successive measurement steps and $\omega(\varphi)=-i \log \Theta(\varphi)$ its phase.  The derivation of the above result, and a more extended discussion of the physical meaning of the winding $w$, are given in the Supplemental Material. 

 While in NH lattices displaying the skin effect the spectral winding counts the number of times the energy spectrum in complex plane rotates around a base energy, here 
 the winding $w$ is of dynamical origin and can be regarded as the manifestation of a new geometric phase \cite{r34}.  The dynamical winding $w$ turns out to be equal the number of non-degenerate eigenstates of $\hat{U}$, i.e. with distinct Floquet exponents $\mu_l=\exp(-i E_l \tau)$ ($E_l$ are the $L$ eigenenergies of $\hat{H}$) that are visited by the dynamical state in the closed cycle \cite{r38}.  In the absence of degeneracies and dark eigenstates at node $p=q$, one then obtains $\langle m \rangle=w=L$, i.e. 
the mean survival time exactly equals the number of nodes $L$ of the network. We remark that the dynamical winding $w$ is independent of  the specific form of the Hamiltonian $\hat{H}$ and, unlike spectral winding in NH lattices displaying the skin effect, it can be introduced for finite-sized systems. Additionally, owing to the topological origin of $w$, any deformation of the Hamiltonian, that does not introduce degeneracies, cannot change its value, indicating robustness of the mean survival time against e.g. disorder or defects in the system. \\
\par
{\em Photonic mesh lattices with repeated projective measurements: model.}  
Temporal dynamics of optical pulses in fiber coupled loops \cite{r39} has provided over the last few years a fascinating platform to investigate NH phenomena in photonics (see e.g. \cite{r32,r40,r41} and references therein). Here we show that light dynamics in engineered synthetic mesh lattices displays dynamical NH topological features, thus providing an experimentally-accessible setting for the observation of the new kind of geometric phase introduced in Ref.\cite{r34}. The system  consists of two fiber loops of slightly different lengths $L \pm \Delta L$ ($\Delta L \ll L$), the short loop and the long loop, that are connected by a variable directional coupler OC$_1$ with a time-varying coupling angle $\beta=\beta(t)$ [Fig.1(a)]. Light travel times in the two fiber loops are $T_1=T-\Delta T$ (in the short loop) and $T_2=T+\Delta T$ (in the long loop), where $T=L/v_g$, $\Delta T=\Delta L/v_g$, and $v_g$ is the group velocity of the optical pulses in the fiber. A single optical pulse of duration $ \tau < \Delta T$, generated from a laser source (LS), is initially injected into the system via a second variable directional coupler OC$_2$. The coupling angle $\sigma=\sigma(t)$ of OC$_2$ can take only the two values $0$ or $\pi/2$, and is suitably switched between them  to periodically eject a light pulse from the short loop, mimicking a repeated projective quantum measurement, as shown in Fig.1(b). Namely, when $\sigma=0$ the short loop is closed and light does to escape from the loop, whereas when $\sigma=\pi/2$ the short loop is open, the light pulse is ejected from the short loop and its intensity is measured by a photo-detector (PD), as shown in Fig.1(b).\\ 
 \begin{figure}[h]
 \centering
    \includegraphics[width=0.45\textwidth]{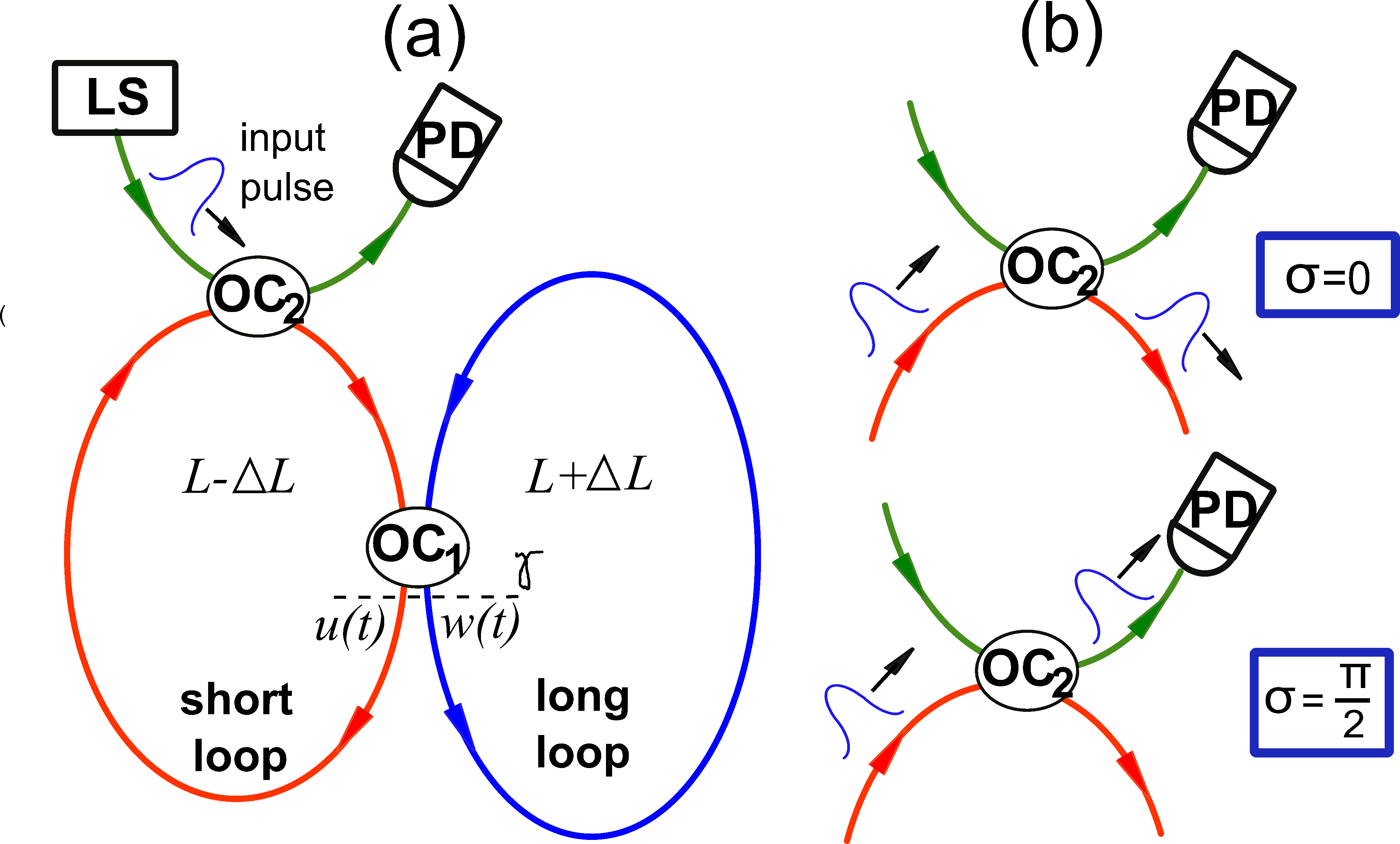}
  \caption{ \small (a) Schematic of the coupled-fiber loop setup that realizes a synthetic mesh lattice via time multiplexing of optical pulses circulating in the loops. Two fiber loops of slightly different lengths $L \pm \Delta L$, the short and the long loops, are coupled by a variable directional coupler OC$_1$ with a time-controlled coupling angle $\beta=\beta(t)$. An input pulse, generated by a laser source (LS), is initially injected into the short loop via a second variable optical coupler OC$_2$. (b) Schematic of the measurement scheme. The coupling angle $\sigma$ of OC$_2$ can take only the two values $\sigma=0$ or $\sigma= \pi/2$. For $\sigma=0$, the short fiber loop is closed, whereas for $\sigma= \pi/2$ the short fiber loop is open, light is ejected from the loop and its intensity monitored by a photo-detector (PD).}
\end{figure}
Let us indicate by $u(t)$ and $w(t)$  the light pulse amplitudes in the short and long loops, respectively, at the output port $\gamma$ of the optical coupler OC$_1$ [see Fig.1(a)]. The synthetic mesh lattice is realized by time multiplexing, i.e. by sampling the pulse amplitudes $u(t)$ and $w(t)$ at discretized times 
\begin{equation}
t=t_n^{(m)}=mT+ n \Delta T, 
\end{equation}
where $n=0, \pm 1, \pm2 ,...$ is the time slot index, each of duration $\Delta T$ and analogous to discrete spatial lattice sites of Hamiltonian $\hat{H}_{eff}$, and $m=1,2,3,....$ is the discrete time step describing the dynamical evolution of the system. The coupling angle $\beta=\beta(t)$ is driven such that  $\beta(t)$  is a periodic function of time with period $T$, so that $\beta(t_n^{(m)})$ is independent of time step $m$.  After letting $\beta_n=\beta(t_n^{(m)})$, $u_n^{(m)}=u(t_{n+1}^{(m)})$, and $w_n^{(m)}=w(t_n^{(m)})$, without detection ($\sigma=0$) the pulse dynamics at successive time steps $m$ is unitary and described by the following set of coupled difference equations (technical details are given in the Supplementary Material)
\begin{eqnarray}
u_n^{(m)} & = \cos \beta_{n+1} u_{n+1}^{(m-1)}+i \sin \beta_{n+1} w_{n}^{(m-1)} \\
w_n^{(m)} & = i \sin \beta_{n} u_{n}^{(m-1)}+ \cos \beta_{n} w_{n-1}^{(m-1)}.
\end{eqnarray}
 \begin{figure}[h]
 \centering
    \includegraphics[width=0.49\textwidth]{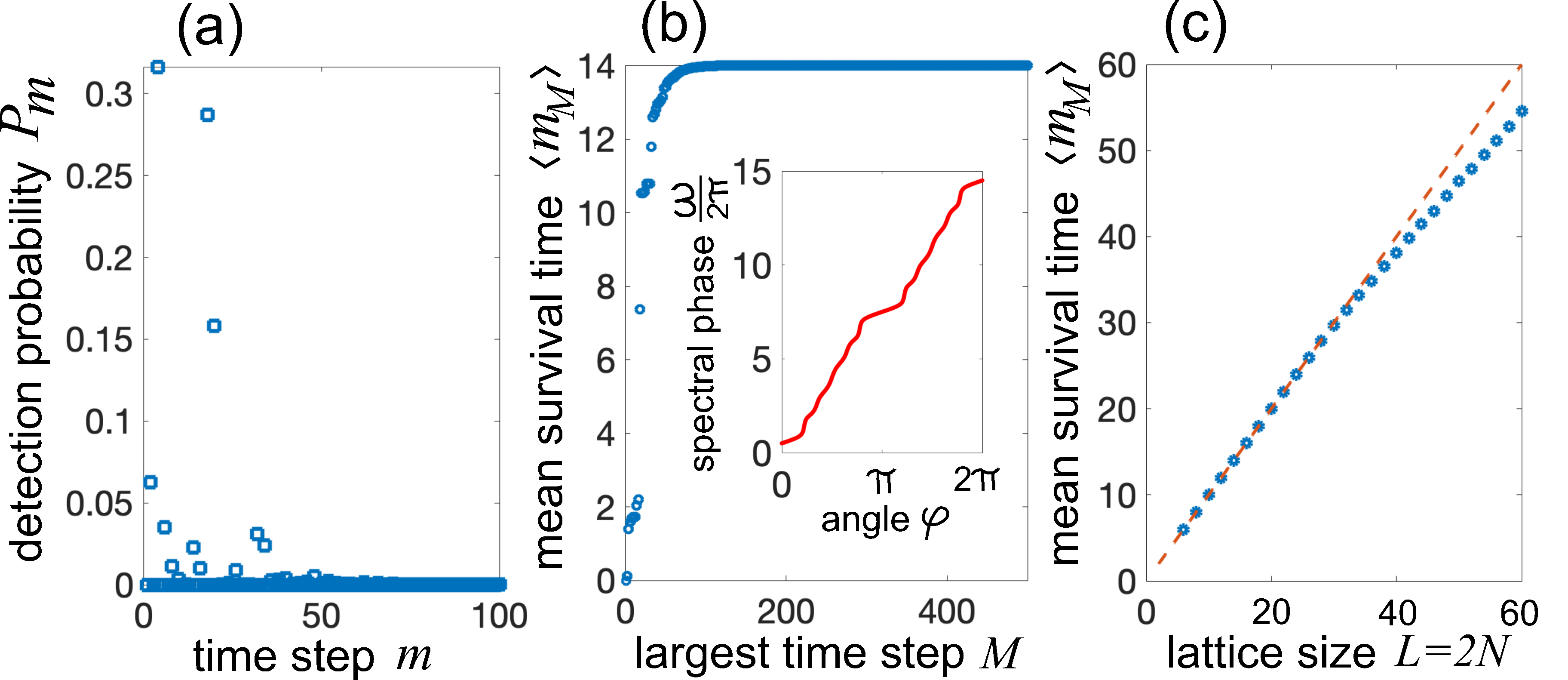}
  \caption{ 	\small Dynamical topological winding in the photonic mesh lattice with uniform coupling angles $\beta= \pi/6$ and size $L=2N=14$.  (a) Distribution of the detection probability $P_m$ versus propagation step $m$ for initial pulse excitation at site $q=2$ and detection at $p=q$. (b) Behavior of the mean survival time $\langle m_M \rangle=\sum_{m=1}^
{M} m P_m$ versus the largest observation time step $M$. The inset shows the numerically-computed  behavior of the phase $\omega(\varphi)$ (in units of $ 2 \pi$) versus angle $\varphi$. Note that as $M$ increases the mean survival time $\langle m_M \rangle$  rapidly converges toward the winding $w=(1/ 2 \pi) \int_{0}^{2\pi} d \varphi (d \omega / d \varphi)=(1/ 2 \pi) [  \omega(2 \pi)- \omega(0) ] =L=14$. (c) Behavior of the mean survival time $\langle m_M \rangle$ for $M=M_{max}=500$ and for increasing values of the lattice size $L=2N$. The dashed line corresponds to the predicted behavior of the winding $w=L$. Pulse excitation and detection is at sites $q=p=2$. } 
\end{figure}
\noindent
A finite number $L=2N$ of lattice sites (discrete time slots), indicated by $|l \rangle$, is simply obtained by assuming $\beta_n= \pi/2$ for $n \leq 1$ and $n \geq (N+1)$: in this case the site index $n$ can be restricted to  the range $n=1,2,...,N$. By letting $|\psi_m \rangle=(u_1^{(m)},u_2^{(m)},..,u_N^{(m)},w_1^{(m)},w_2^{(m)},...,w_N^{(m)})^T$, i.e. $\langle l | \psi_m \rangle=u_l^{(m)}$ for $l=1,2,..,N$ and  $\langle l  | \psi_m \rangle=w_{l-N}^{(m)}$ for $l=N+1,N+2,...,2N$,  Eqs.(5) and (6) can be cast in the compact form $| \psi_{m} \rangle=\hat{U} | \psi_
{m-1} \rangle$, where $\hat{U}$ is a unitary $2N \times 2N$ matrix.  For initial single-pulse excitation of the system at time slot (site index) $q$ in the short loop, the above equations should be integrated with the initial condition $ | \psi_1 \rangle= \hat{U} |q \rangle= \cos \beta_q |q-1 \rangle+ i \sin \beta_q |N+q \rangle$.
The non-unitary (dissipative) dynamics, corresponding to period light pulse ejection at site $l=p$ in the short loop, is obtained by periodically switching the coupling angle $\sigma(t)$ from $0$ to $\pi/2$ at time instants $t_p^{(m)}$ for a short interval $\sim \Delta T$ (see Supplementary Material for details). Such a periodic pulse ejection basically corresponds to repetitive projective measurements in the quantum mechanical context, and described by the operator $(1-|p \rangle \langle p|)=(1- \hat{\Gamma})$. We remark that in our setting we deal with classical optical fields and use mean-field equations,  however the same scenario would be observed if the system is excited by a single photon pulse, the detector 'click' corresponding indeed to a projective quantum measurement and annihilation of the photon. The advantages of using classical optical fields (rather than single photons) are: (i) the simplicity of excitation and detection apparatus, and (ii) single-run measurements, avoiding repeated experimental runs and the need of statistical averaging, the detection probability distribution $P_m$  simply being obtained from the intensity measurements at the photo-detector in a single run. \\
\\
{\em Dynamical topological winding in photonic mesh lattices.}
To illustrate the topological quantization of the mean survival time in the system, let us consider as an example a homogenous lattice, with $\beta_n= \beta$ constant and different than $\pi/2$ for $n=2,3,...,N$. System evolution is monitored up to the maximum time step $m=M_{max}=500$, which is experimentally feasible \cite{r32,r40,r41,r42}. Figure 2(a) shows the numerically-computed behavior of the detection probability distribution $P_m=|u_{p}^{(m)}|^2$ in a lattice of size $L=2N=14$ and $\beta=\pi/6$, assuming $p=q=2$ for the excitation and detection nodes, whereas Fig.2(b) depicts the behavior of the mean survival time $\langle m_M  \rangle \equiv \sum_{m=1}^{M} m P_m$, up to the observation step $M$, as $M$ is increased toward $M_{max}$. As it can be seen, $\langle m_M \rangle$ rapidly converges to the integer $w=L=14$, which is the number of nodes of the lattice. This is in agreement with the circumstance that the eigenvalues of $\hat{U}$ are not degenerate and the corresponding eigenstates are not dark at site $p=q=2$. The inset in Fig.2(b) shows to the numerically-computed behavior of the phase $ \omega(\varphi)$, which is computed from the spectral analysis described in the Supplementary Material.
For a given largest propagation step $M=M_{max}$, the setup can correctly measure the winding $w$ for a system size $L$ (nodes of the network) not too large, i.e. there is an upper limit to the Hilbert space dimension set by the largest possible observation time $M=M_{max}$ of the system. The reason thereof is that, according to the time-energy uncertainty principle, as the system size $L$ increases the energy level spacing decreases, and thus a longer observation time is required to resolve the spectral features of the Hamiltonian. As an example, Fig. 2(c) shows the behavior of the mean survival time $\langle m_{M_{max}} \rangle$ for increasing values of the system size $L$, indicating that for the largest propagation step $M_{max}=500$ the winding $w$ is correctly measured for a system size up to $L \sim 36$. Finally, it should be mentioned that the convergence of $\langle m_M \rangle$ toward the topological winding $\langle m \rangle= w$ can become extremely slow  near quasi-energy degeneracies of $\hat{U}$, where the variance of the distribution $P_m$ takes extremely large values and  intermittency is observed. Such a circumstance typically occurs when $\hat{H}$ displays quasi-degenerate eigenenergies $E_n \simeq E_l$,  near resonances such that $E_n \tau \simeq E_l \tau$ (mod. $ 2 \pi$), or for $\tau \rightarrow 0$ owing to Zeno effect. Such limitations have been discussed at length in Ref. \cite{r38}. In such regimes, the experimental observation of the topological winding can be challenging, requiring extremely long propagation times. 
\par
Since the winding $w$ is of topological nature, it does not vary as we deform the Hamiltonian, i.e. change the coupling angles $\beta_n$ from the homogeneous case, even in a disordered manner, provided that quasi-energy degeneracies are avoided. As an illustrative example, let us consider the same case of Fig.2 (homogeneous coupling angles), except than we introduce a defect by letting $\beta_n=\tilde{\beta}$ for $n=n_0$. Figure 3 shows the numerically-computed behavior of the mean survival time $\langle m_{M} \rangle$, up to the propagation step $M=M_{max}=500$, for $n_0=4$ as $\tilde{\beta}$ is increased from $\beta= \pi/6$ (homogeneous limit) to $\tilde{\beta}= \pi/2$. The lattice size is fixed at $L=2N=14$; pulse excitation and detection is at site $q=p=2$ as in Fig.2. Note that the value of $\langle m_{M} \rangle$ remains unchanged and locked at the value $w=2L=14$ for a wide range of $\tilde{\beta}$, with a rather abrupt change to $\langle m_{M} \rangle \simeq 6$ as $\tilde{\beta}$ approaches the edge value $\tilde{\beta}= \pi/2$.
The reason of such an abrupt change stems from the circumstance that, at  $\tilde{\beta}= \pi/2$, the network splits into two decoupled sublattices, and the effective dynamics occurs in one sublattice solely with reduced dimensionality $L_r=2(n_0-1)=6$.\\
    
   \begin{figure}[h]
 \centering
    \includegraphics[width=0.45\textwidth]{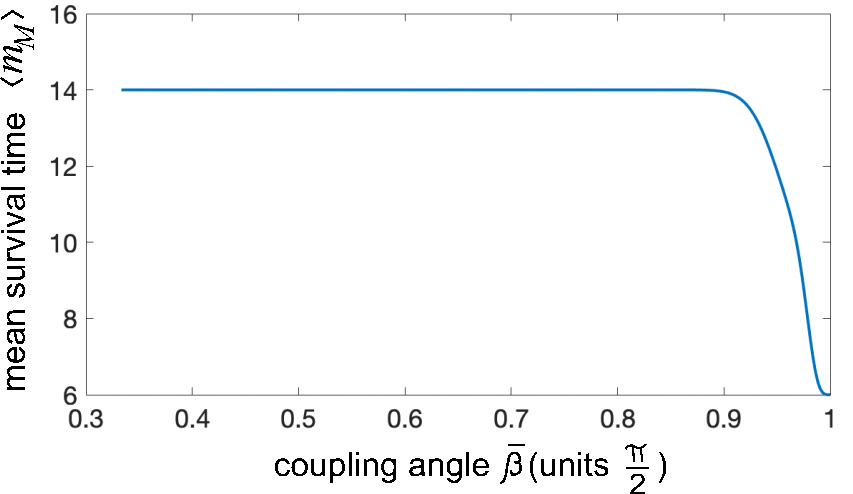}
  \caption{ \small Dynamical topological winding in the photonic mesh lattice with defect. The panel shows the behavior of  the mean survival time $\langle m_M \rangle=\sum_{m=1}^
{M} m P_m$ for $M=M_{max}=500$ versus $\bar{\beta}$ in a lattice of size $L=2N=14$. Other parameter values are given in the text. Note that $\langle m_M \rangle$ does not change as $\bar{\beta}$ is varied and remains locked at the value of the winding number $w=2L=14$, except for a narrow interval near $\bar{\beta}= \pi/2$ where the effective dimensionality of the network is decreased down to $L_r=6$.} 
\end{figure}

\noindent

{\em Conclusions.} We have shown that dynamical non-Hermitian topological windings can be observed in photonic lattices as a manifestation of a new type of geometric phase recently introduced in Ref. \cite{r34} for the problem of mean return time in quantum systems. Such a geometric phase arises due to repeated projective measurements in the unitary evolution of a quantum system, which results in an effective cyclic dissipative dynamics. Here we pointed out that such a geometric phase can be unravelled in dynamical evolution of classical systems as well, such as in the dynamics of optical pulses in coupled fiber loops. Owing to its topological origin, the winding number is quantized and remains unchanged under Hamiltonian deformations that avoid energy level crossing and do not introduce dark states. Our results indicate that photonics could provide an accessible platform for the experimental demonstration of dynamical NH topological windings \cite{r34}, whose observation remains challenging in truly quantum systems. Since the dynamical convergence toward the topological winding becomes critical near system degeneracies or resonances \cite{r38}, probing NH dynamical windings and mapping the probability distribution $P_m$ provide interesting tools to detect system degeneracies, which could be of potential relevance in advanced sensing applications.\\
\\
\noindent
{\bf Disclosures}. The author declares no conflicts of interest.\\
\\
{\bf Data availability}. No data were generated or analyzed in the presented research.\\
\\
{\bf Funding}. Agencia Estatal de Investigacion (MDM-2017-0711).\\
\\
{\bf Supplemental document}. See Supplement 1 for supporting content.

\newpage

%%%%%%%%%%%%%%%%%%%%%%%%%%%%%%%
% References with full titles %
%%%%%%%%%%%%%%%%%%%%%%%%%%%%%%%

%\footnotesize
 {\bf References with full titles}\\
 \\
 \noindent
1. M. Z. Hasan and C. L. Kane,  Colloquium: Topological insulators,
Rev. Mod. Phys. {\bf 82}, 3045 (2010).\\
2.  X.-L. Qi and S.-C. Zhang, Topological insulators and superconductors,
Rev. Mod. Phys. {\bf 83}, 1057 (2011).\\
3. L. Lu, J.D. Joannopoulos, and M. Solja\u{c}i\'{c}, Topological photonics,
Nature Photon. {\bf 8}, 821 (2014).\\
4. T. Ozawa, H.M. Price, A. Amo, N. Goldman, M. Hafezi, L. Lu, M.C. Rechtsman, D. Schuster, J. Simon, O. Zilberberg, and I. Carusotto Topological photonics, Rev. Mod. Phys. {\bf 91}, 015006 (2019).
5. A.B. Khanikaev and A. Al\'u, 
Topological photonics: robustness and beyond,
Nature Commun. {\bf 15}, 931 (2024).\\
6. A. Szameit and M.C. Rechtsman, Discrete nonlinear topological photonics, Nat. Phys. (2024). https://doi.org/10.1038/s41567-024-02454-8.\\
7. Z. Gong, Y. Ashida, K. Kawabata, K. Takasan, S. Higashikawa, and M. Ueda, 
Topological Phases of Non-Hermitian Systems, 
Phys. Rev. X {\bf 8}, 031079 (2018).\\
8. C.-H. Lee and R. Thomale, Anatomy of skin modes and topology in non-Hermitian systems,
Phys. Rev. B {\bf 99}, 201103 (2019).\\
9. Y. Ashida, Z. Gong, and M. Ueda, Non-Hermitian Physics, Adv. Phys. {\bf 69}, 249 (2020).\\
10. K. Kawabata, K. Shiozaki, M. Ueda and M. Sato, Symmetry and Topology in Non-Hermitian Physics, Phys. Rev. X {\bf 9}, 041015 (2019).\\
11. N. Okuma, K. Kawabata, K. Shiozaki, and M. Sato, Topological Origin of Non-Hermitian Skin Effects,
Phys. Rev. Lett. {\bf 124}, 086801 (2020).\\
12. Z. Yang, K. Zhang, C. Fang, and J. Hu, Non-Hermitian Bulk-Boundary Correspondence and Auxiliary Generalized Brillouin Zone Theory,
Phys. Rev. Lett. {\bf 125}, 226402 (2020).\\
13. E.J. Bergholtz, J.C. Budich, and F.K. Kunst, Exceptional topology of non-Hermitian systems,
Rev. Mod. Phys. {\bf 93}, 015005 (2021).\\
14. K. Wang, A. Dutt, K.Y. Yang, C.C. Wojcik, J. Vuckovic, and S. Fan, Generating arbitrary topological windings of a non-Hermitian band, Science {\bf 371}, 1240 (2021).\\
15. K. Wang, A. Dutt, C.C. Wojcik,
and S. Fan,  Topological complex-energy braiding of
non-Hermitian bands, Nature {\bf 598}, 59 (2021).\\
16. K. Ding, C. Fang, and G. Ma, Non-Hermitian topology and exceptional-point geometries, 
Nature Rev. Phys. {\bf 4}, 745 (2022).\\
17. A. Banerjee, R. Sarkar, S. Dey, and A. Narayan, Non-Hermitian topological phases: principles and prospects, J. Phys.: Condens. Matter {\bf 35}, 333001 (2023).\\
18. R. Lin, T. Tai, M. Yang, L. Li,  and C.H Lee,
Topological Non-Hermitian skin effect, Front. Phys. {\bf 18}, 53605 (2023).\\
19. T. Tai and C.-H. Lee,  Zoology of non-Hermitian spectra and their graph topology,
Phys. Rev. B {\bf 107}, L220301 (2023).\\
20. M. S. Rudner and L. S. Levitov, Topological Transition in a Non-Hermitian Quantum Walk,
Phys. Rev. Lett. {\bf 102}, 065703 (2009).\\
21.J. M. Zeuner, M. C. Rechtsman, Y. Plotnik, Y. Lumer,
S. Nolte, M. S. S. Rudner, M. Segev, and A. Szameit,
Observation of a Topological Transition in the Bulk of a
Non-Hermitian System, Phys. Rev. Lett. {\bf 115}, 040402
(2015).\\
22. T. Rakovszky, J.K. Asboth, and A. Alberti, Detecting topological invariants in chiral symmetric insulators via losses,
Phys. Rev. B {\bf 95}, 201407 (2017).\\
23. X. Zhan, L. Xiao, Z. Bian, K. Wang, X. Qiu, B.C. Sanders, W. Yi, and P. Xue, 
Detecting Topological Invariants in Nonunitary Discrete-Time Quantum Walks,
Phys. Rev. Lett. {\bf 119}, 130501 (2017).\\
24. F. Cardano, A. D'Errico, A. Dauphin, M. Maffei, B. Piccirillo, C. de Lisio, G. De Filippis, V. Cataudella, E. Santamato, L. Marrucci, M. Lewenstein, and P. Massignan, Detection of Zak phases and topological invariants in a chiral quantum walk of twisted photons, Nat. Commun. {\bf 8}, 15516 (2017).\\
25. L. Xiao, X. Qiu, K. Wang, Z. Bian, X. Zhan, H. Obuse, B.C. Sanders, W. Yi, and P. Xue,
Higher winding number in a nonunitary photonic quantum walk,
Phys. Rev. A {\bf 98}, 063847 (2018).\\
26. S. Longhi, Probing one-dimensional topological phases in waveguide lattices with broken chiral symmetry, Opt. Lett. {\bf 43}, 4639 (2018).\\
27. A. D'Errico, F. Cardano, M. Maffei, A. Dauphin, R. Barboza, C. Esposito, B. Piccirillo, M. Lewenstein, P. Massignan, and L. Marrucci,
Two-dimensional topological quantum walks in the momentum space of structured light, Optica {\bf 7}, 108 (2020).\\ 
28. B. Zhu, Y. Ke, H. Zhong, and C. Lee,
 Dynamic winding number for exploring band topology, 
Phys. Rev. Research 
{\bf 2}, 023043 (2020).\\
29. K. Wang, T. Li, L. Xiao, Y. Han, W. Yi, and P. Xue, 
Detecting Non-Bloch Topological Invariants in Quantum Dynamics,
Phys. Rev. Lett. {\bf 127}, 270602 (2021).\\
30. S. Longhi, Topological Phase Transition in non-Hermitian Quasicrystals, Phys. Rev. Lett. {\bf 122}, 237601 (2019).\\
31. J. Claes and T.L. Hughes,  Skin effect and winding number in disordered non-Hermitian systems,
Phys. Rev. B {\bf 103}, L140201 (2021).\\
32. S. Weidemann, M. Kremer, S. Longhi and A. Szameit,
Topological triple phase transition in non-Hermitian Floquet quasicrystals, Nature {\bf 601}, 354 (2022).\\
33. Q. Lin, T. Li, L. Xiao, K. Wang, W. Yi, and P. Xue, Topological Phase Transitions and Mobility Edges in Non-Hermitian Quasicrystals,
Phys. Rev. Lett. {\bf 129}, 113601 (2022).\\
34. F. A. Gr\"unbaum, L. Vel\'azquez, A. H.Werner, and R.F. Werner,  Recurrence for Discrete Time Unitary Evolutions,
Commun. Math. Phys. {\bf 320}, 543 (2013).\\
35. S. Dhar, S. Dasgupta, A. Dhar, and D. Sen,
Detection of a quantum particle on a lattice under repeated projective measurements, Phys. Rev. A {\bf 91}, 062115 (2015).\\
36. H. Friedman, D.A. Kessler, and E. Barkai, Quantum walks: The first detected passage time problem, Phys. Rev. E {\bf 95}, 032141 (2017).\\
37. F. Thiel, E. Barkai, and D.A. Kessler,
First Detected Arrival of a Quantum Walker on an Infinite Line, Phys. Rev. Lett. {\bf 120}, 040502 (2018).\\
38. R. Yin, K. Ziegler, F. Thiel, and E. Barkai,
Large fluctuations of the first detected quantum return time, Phys. Rev. Research {\bf 1}, 033086 (2019).\\
39. S. Longhi, Nonexponential Decay Via Tunneling in Tight-Binding Lattices and the Optical Zeno Effect,
Phys. Rev. Lett. {\bf 97}, 110402 (2006).\\
40. P. Biagioni, G. Della Valle, M. Ornigotti, M. Finazzi, L. Du\`o, P. Laporta, and S. Longhi, Experimental demonstration of the optical Zeno effect by scanning tunneling optical microscopy, Opt. Express {\bf 16},  3762 (2008).\\
41. F. Dreisow, A. Szameit, M. Heinrich, T. Pertsch, S. Nolte, A. T\"unnermann, and S. Longhi,
Decay Control via Discrete-to-Continuum Coupling Modulation in an Optical Waveguide System, Phys. Rev. Lett. {\bf 101}, 143602 (2008).\\
42.  A. Regensburger, C. Bersch, B. Hinrichs, G. Onishchukov, A. Schreiber, C. Silberhorn, and U. Peschel, Photon propagation in a discrete fiber network: an interplay of coherence and losses, Phys. Rev. Lett. {\bf 107}, 233902 (2011).\\
43. A. Regensburger, C. Bersch, M.-A. Miri, G. Onishchukov, D.N. Christodoulides, and U. Peschel,
 Parity-time synthetic photonic lattices, Nature {\bf 488}, 167 (2012).\\
44. M. Wimmer, A. Regensburger, M.-A. Miri, C. Bersch, D.N. Christodoulides, and U. Peschel, Observation of optical solitons in PT-symmetric lattices, Nat. Commun. {\bf 6}, 7782 (2015).\\
45. S. Wang, C. Qin, W. Liu, B. Wang, F. Zhou, H. Ye, L. Zhao, J. Dong, X. Zhang, S. Longhi, and P. Lu, High-order dynamic localization and tunable temporal cloaking in ac-electric-field driven synthetic lattices, Nature Commun. {\bf 13}, 7653 (2022).\\

\end{document}